# A Novel Detection Algorithm Efficient for Turbo coded CDMA Signals in Detect and Forward Cooperative Channels

Ebrahim Karami

Centre for Wireless Communications (CWC), University of Oulu,

P.O. Box 4500, FIN-90014, Oulu, Finland

*Abstract—* In this paper, a new detection algorithm is proposed for turbo coded Code Division Multiple Access (CDMA) signals in detect and forward cooperative channels. Use of user cooperation makes much improvement in the performance of CDMA systems. Due to the special structure of CDMA systems, cooperative schemes increase the sum and cutoff capacities of CDMA based wireless systems and improve the quality of user-partner link which enhances the overall performance of the system. In this paper, a new combining scheme is proposed that makes the receiver more robust against the decision errors in the partner link. This structure is simulated for punctured ½ rate 4 states turbo code in a channel with first order Markov time variation and different Rice factor variances. Through various simulations, it is shown when the channel estimates are available in the partner and receiver, the cooperation between users provides much diversity gain especially while using the new proposed combining algorithm.

## I. INTRODUCTION

In a wireless channel, fading is the most important limiting factor for mobility and consequently sum and cutoff capacity of mobile users. To combat the fading effects, diversity techniques are proposed. Diversity is to receive multiple copies. Diversity techniques can be classified into transmit [1], receive

[2], time [3], frequency [4], code [5], space and polarization diversity [6]. Except for space diversity, the other techniques provide diversity gain at the cost of reduction in sum capacity. In the recent years, antenna arrays and their extension as Multiple-Input Multiple-Output (MIMO) systems have gained interest in improving capacity and mobility in mobile wireless channels under severe fading effects [7]. Due to the physical limitation in the handsets, use of antenna array in them is impractical, especially in lower frequency bands corresponding to long wavelengths. According to the Lee model [8] for wireless fading channels, the minimum required space between elements of an array in a handset is $0.2\lambda$ where $\lambda$ is the wavelength of carrier that is about 6.7cm and 3.1cm in the second and third generations of wireless standards, respectively. Therefore, application of antenna arrays is restricted to base stations where size-limitation is less important than in the handset.

Use of cooperation between users in the transmission process is an efficient solution for the above problem [9]. In a cooperative wireless communication, unlike the usual antenna arrays, the link between elements of antenna array is wireless. Therefore, the cooperative channels are so called virtual arrays or distributed antenna array systems. In common antenna arrays because of wired links between the array elements, the problems related to attenuation and bandwidth of these links don't exist and therefore the system can be designed with simplicity. In the cooperative systems, the quality of user-partner link has a great influence on the performance of the system [10, 11]. There are several cooperation signaling methods such as detect and forward, amplify and forward, and coded cooperation [12]. In detect and forward cooperative channels which is very close to traditional relaying idea, data received by partner is detected and then its modulated again and transmitted toward main receiver. Therefore this new transmitted data suffers from decisions errors occurred in the partner link.

The use of CDMA as an access technique not only solves the above problem [13], but also eliminates the need for additional wireless channel for user-partner link. On the other hand, while using CDMA, the requirement of precise timing between user and partner transmission time-slots is circumvented. In all existing researches on the cooperative systems, especially for detect and forward schemes, in the design of the receiver, the decision errors occurring in partner link are ignored. Using turbo codes in cooperative systems is proposed to approach the capacity achievable by relay based channels [14-16]. In this paper, in addition to simulation of the turbo coded cooperative based multiuser CDMA; for time varying channels by considering the effect of partner selection set size and Ricean factor of users in cell, a new combining scheme is proposed to make the performance of the receiver more robust to the decision errors occurring in partner link.

The rest of this paper is organized as follows. In Section II, the models used for signal, spreading, and time varying channel are introduced. In Section III, the receiver structures with different combining schemes are introduced. In Section IV, the simulation results for the proposed receiver are presented, and the concluding remarks are presented in Section V.

## II. SYSTEM MODEL

**A. Signal Model**

Assume $b_k$ as data symbol coming from data source at time index *k*. As shown in Fig. 1, each user encodes data symbols $b_k$ by a punctured turbo encoder with rate of ½ as shown in Fig. 2 and then its corresponding output symbol, $\boldsymbol{d}_k$, that has two bits $d_{2k-1}$ and $d_{2k}$ as its elements are spread by the spreading sequence $\boldsymbol{s}_1$ which is a $1 \times N_c$ vector as follows

$$\boldsymbol{t}_{2k-i} = \boldsymbol{s}_1 d_{2k-i}, \quad i = 0,1, \tag{1}$$

where $\boldsymbol{t}_{2k-i}$ is $1 \times N_c$ transmitted vector in time index *k*.

As illustrated in Fig. 3, a copy of this signal is received by a pre-selected partner as follows,

$$\boldsymbol{r}_{p,2k-i} = h_{in,k} \boldsymbol{t}_{2k-i} + \boldsymbol{I}_{p,2k-i} + \boldsymbol{w}_{p,2k-i}, \quad i = 0,1, \tag{2}$$

where $r_{p,2k-i}$ is the received vector by the partner, $I_{p,2k-i}$ and $w_{p,2k-i}$ are multiple access interference and AWGN noise, respectively, in the front end of the partner receiver, and $h_{in,k}$ is the user-partner coefficient that is scalar due to single-antenna assumption for users. The received vector is despreaded by the following equation and decoded by the turbo decoder. The block diagram of the turbo decoder is shown in Fig. 4.

$$y_{p,2k-i} = s_1^H r_{p,2k-i}, \quad i=0,1, \tag{3}$$

where the superscript $H$ is the conjugate transpose operator and $y_{p,2k-i}$ is the output of the despreader. For each user, a partner selection set is defined. The size of this set depends on the required privacy degree. The members of this set are so called the friends of user and have full access to the data of user. Of course to overcome this limitation, encryption algorithms can be considered but this issue is not of interest and thus not regarded in the scope of this paper. The partner is chosen to have the best quality propagation environment to the user. Partner can be selected by many criteria the most common of which is the nearest user in cell [18]. In this paper, as a result of the time varying nature of the channel due to the movement of the users, the partner is selected as such to have the largest LoS component. Therefore, the user-partner link is usually Ricean with high enough SNR and Ricean factor. Thus, the data detected by the partner is very close to the original transmitted coded data. The detected data, $\hat{d}_k$, that has two bits $\hat{d}_{2k-1}$ and $\hat{d}_{2k}$ as its elements is then spread by the spreading sequence $s_2$ and retransmitted to the receiver as follows,

$$t_{p,2k-i} = s_2 \hat{d}_{2k-i}, \quad i=0,1, \tag{4}$$

where $t_{p,2k-i}$ is $1 \times N_c$ vector transmitted by the partner. Of course, in this section only the detection of the coded transmitted signal is performed and to keep the synchronization of user and partner the decoding and re-encoding of the coded signal is ignored. Consequently, the lack of synchronization observed between user and its partner can be negligible and ignored to reduce the complexity in the design of the combining algorithm at the receiver side.

In the receiver side, the sum of two copies of the signal transmitted by user and its partner, interference vector $I_{2k-i}$ due to the other users and their partners using the same frequency band, and AWGN noise $w_k$ is received in $N$ elements of receiver array as follows,

$$r_{2k-i} = h_{u,k} t_{2k-i} + h_{p,k} t_{p,2k-i} + I_{2k-i} + w_{2k-i}, \quad i = 0,1, \tag{5}$$

where $h_{u,k}$ and $h_{p,k}$ are $N \times 1$ vectors of channel coefficients for user-receiver and partner-receiver, respectively, where $N$ is the number of the receiver array elements. Therefore $r_{2k-i}$, $I_{2k-i}$, and $w_{2k-i}$ are $N \times N_c$ matrices.

**B. Time Variation Model of the Channel**

Due to the short distance between user and partner, the user-partner link is a Ricean channel that is a combination of a constant part and a time varying part as the following equation.

$$h_{in,k} = h_{in,0} + \tilde{h}_{in,k}, \tag{6}$$

where $h_{in,0}$ is the nearly constant part which can be due to the LoS components that are fairly constant in a long block of the data, and $\tilde{h}_{in,k}$ is the time varying part which is mainly due to the No-LoS (NLoS) components. Because of a relatively long distance between user and its partner to base station the corresponding channel coefficients, same as time-varying part, are severely variable in the duration of data block transmission with the autocorrelation that follows the following equation [18]

$$\begin{aligned} E\left\{ \tilde{h}_{in,k} \left[\tilde{h}_{in,l}\right]^* \right\} &\cong J_0\left(2\pi f_{in,D} T |k-l|\right), \\ E\left\{ h_{m,k}^i \left[h_{m,l}^i\right]^* \right\} &\cong J_0\left(2\pi f_{m,D}^i T |k-l|\right), \quad m = u, p \end{aligned} \tag{7}$$

where $h_{u,k}^i$ and $h_{p,k}^i$ are $i$th elements of $\mathbf{h}_{u,k}$ and $\mathbf{h}_{p,k}$ or in other word the channel coefficients between user and its partner, respectively, to $i$th elements in base station receiver, $J_0(.)$ is the zero-order Bessel function of the first kind, superscript * denotes the complex conjugate, $f_{in,D}$ is the Doppler frequency shift between transmitter and partner, $f_{u,D}^i$ and $f_{p,D}^i$ are Doppler frequency shift between transmitter and its partner, respectively, to $i$th elements in base station receiver, and $T$ is the duration of each symbol. According to the Wide Sense Stationary Uncorrelated Scattering (WSSUS) model of Bello [19], all the channel taps are independent, namely, $h_{u,k}$ and all $h_{u,k}^i$s and $h_{p,k}^i$s vary independently, according to the autocorrelation model of (7). The normalized spectrum for each tap $\tilde{h}_{in,k}$ is

$$S_{in,k}(f) = \begin{cases} \dfrac{1}{\pi f_{in,D} T \sqrt{1 - \left(\dfrac{f}{f_{in,D}}\right)^2}}, & |f| < f_{in,D} T \\ 0, & otherwise. \end{cases} \tag{8}$$

The exact modeling of the vector process $\tilde{h}_{in,k}$ with a finite length Auto-Regressive (AR) model is impossible. For implementation of a channel estimator, MIMO channel variations $\tilde{h}_{in,k}$ and $h_{u,k}^i$s and $h_{p,k}^i$s can be approximated by the following AR process of order L

$$\begin{aligned}\tilde{h}_{in,k} &= \sum_{l=1}^{L} \alpha_{in,l} \tilde{h}_{in,k-l} + v_{in,k}, \\ h_{m,k}^i &= \sum_{l=1}^{L} \alpha_{m,l}^i h_{m,k-l}^i + v_{m,k}^i, \quad m = u, p \text{ and } i = 1,2,...,N\end{aligned} \tag{9}$$

where $\alpha_{in,l}$, $\alpha_{u,l}^i$s, and $\alpha_{p,l}^i$s are the $l$th coefficients in AR models of $h_{in,k}$, $h_{u,k}^i$s and $h_{p,k}^i$s and $v_{in,k}$, $v_{u,k}^i$s, and $v_{p,k}^i$s are zero-mean i.i.d. complex Gaussian processes with variances given by

$$E\left(v_{m,k}^i \left[v_{m,k}^i\right]^*\right) = \sigma_{v_{m,k}^i}^2, \quad m = u, p, in \quad and \quad i = 1, 2, ..., N. \tag{10}$$

For the optimum selection of channel AR model parameters from correlation functions, Wiener equations can be used applying the L following equations

$$J_0\left(2\pi f_{m,D} T |k-t|\right) = \sum_{l=1}^{L} J_0\left(2\pi f_{m,D} T |k-l-t|\right) \alpha_{m,l},$$

$$m = u, p, i, \quad i = 1, 2, ..., N \quad and \quad t = k - L, k - L + 1, ..., k - 1. \tag{11}$$

The length of the channel model must be chosen to a minimum of 90% of the energy spectrum of each channel coefficients contained in the frequency range of $|f| < f_{m,D} T$.

The speed of channel variations is dependent on the Doppler shift, or equivalently on the relative velocity between the receivers and the transmitter elements. A reasonable assumption, which is conventional in most of the scenarios, is the equal Doppler shifts $f_{u,D}^i = f_{u,D}$ and $f_{p,D}^i = f_{p,D}$ for the elements of the receiver array. Therefore, channel coefficients for user and its partner to the array of the base station can be rewritten in vector forms as the following equation

$$\mathbf{h}_{m,k} = \sum_{l=1}^{L} \alpha_{m,l} \mathbf{h}_{m,k-l} + \mathbf{v}_{m,k}, \quad m = u, p, \tag{12}$$

where $\mathbf{h}_{u,k}$ and $\mathbf{h}_{p,k}$ are the channel vectors and $\alpha_{u,l}$s and $\alpha_{p,l}$ are the coefficients of order $L$ of the AR channel model variations that can be calculated from (4) which for first order channel model, by solving Wiener equation, the values of the $\alpha_u$ and $\alpha_p$ are calculated as

$$\alpha_m = J_0(2\pi f_{m,D}T), \quad m = u, p. \tag{13}$$

It is obvious that the larger Doppler rates leads to the smaller $\alpha$ and, therefore, faster channel variations.

## III. THE RECEIVER DESIGN

The block diagram of the receiver is shown in Fig. 5. As it can be seen, the main part of the receiver is combining block where the signal received directly and indirectly from user and partner, respectively, must be combined. In this paper, 3 combining algorithms are presented to combine the signals received from the user and the partner. The first two algorithms are the maximum ratio combining (MRC) and the log likelihood ratio (LLR) detection that are well-known algorithms in the communications literature, and the third algorithm is constructed by certain modifications in the LLR algorithm.

**A. The MRC Algorithm**

In a MRC combiner the receiver uses a two dimensional Rake receiver as follows. All the signals received by N elements of the receiver array is despread by matched filters corresponding to spreading sequences used in user and partner independently by

$$\mathbf{y}_{j,2k-i} = \mathbf{r}_{2k-i}\mathbf{s}_j^H, \quad i = 0,1, \quad j = 1,2. \tag{14}$$

Then by employing the maximum ratio combiner, the 2N outputs of the matched filters are combined and fed into the turbo decoder.

$$\mathbf{y}_{2k-i} = L_C\left(\mathbf{h}_{u,k}^H \mathbf{y}_{1,2k-i} + \mathbf{h}_{p,k}^H \mathbf{y}_{2,2k-i}\right), \quad i = 0,1, \tag{15}$$

where $\mathbf{y}_{j,2k-i}$ are the N outputs of the matched filters, with 2 elements for each, $\mathbf{y}_{2k-i}$ is the output of the maximum ratio combiner which is a two elements vector, and $L_C$ is the scaling factor required by turbo decoder which for $N_I$ interfering user is defined as follows,

$$L_C = \frac{2}{\left(\sigma_w^2 + \frac{N_I}{N_c}\right)}. \tag{16}$$

The denominator of (16) denotes the variance of equivalent noise plus interference i.e. $\sigma_e^2$ when the random spreading is used

$$\sigma_e^2 = \sigma_w^2 + \frac{N_I}{N_c}. \tag{17}$$

It must be noted that the propagation vectors $h_{u,k}$ and $h_{p,k}$ are assumed to be known in the receiver. Also, the inter-user link coefficient is assumed to be known for the partner.

**B. The LLR Algorithm**

The LLR of each received symbol can be calculated as follows,

$$y_{2k-i} = \frac{Pr(d_{2k-i}=1|r_{2k-i})}{Pr(d_{2k-i}=-1|r_{2k-i})} \quad i=0,1, \tag{18}$$

where $Pr(.|.)$ presents the conditional probability. By applying the Bays theorem, then

$$y_{2k-i} = \frac{Pr(r_{2k-i}|d_{2k-i}=1)Pr(d_{2k-i}=1)}{Pr(r_{2k-i}|d_{2k-i}=-1)Pr(d_{2k-i}=-1)}. \tag{19}$$

We assume, there is not any initial knowledge about the transmitted symbols, therefore

$$Pr(d_{2k-i}=1) = Pr(d_{2k-i}=-1) = 1/2. \tag{20}$$

Then by applying (1), (4), (5), and (20) in (19) and assuming correct decisions in partner side i.e. $\hat{d}_{2k-i} = d_{2k-i}$, we have,

$$y_{2k-i} = \frac{exp\left(-\frac{1}{2\sigma_{eq}^2}\|\mathbf{r}_{2k-i} - \mathbf{h}_{u,k}\mathbf{s}_1 - \mathbf{h}_{p,k}\mathbf{s}_2\|^2\right)}{exp\left(-\frac{1}{2\sigma_{eq}^2}\|\mathbf{r}_{2k-i} + \mathbf{h}_{u,k}\mathbf{s}_1 + \mathbf{h}_{p,k}\mathbf{s}_2\|^2\right)}. \quad (21)$$

The output of (21) must be applied to the turbo decoder directly and without any scaling.

## C. The Modified LLR Algorithm

In the conventional method for calculation of LLR and MRC methods, we get an assumption on the completely correctness on decisions in partner side. This assumption can be destructive especially when the user-partner link has low quality. Hence, certain modifications in calculation of LLR defined in (18) are introduced as follows,

$$Pr(r_{2k-i}|d_{2k-i} = +1) = Pr(r_{2k-i}|d_{2k-i} = +1, \hat{d}_{2k-i} = +1) Pr(\hat{d}_{2k-i} = +1|d_{2k-i} = +1)$$
$$+ Pr(r_{2k-i}|d_{2k-i} = +1, \hat{d}_{2k-i} = -1) Pr(\hat{d}_{2k-i} = -1|d_{2k-i} = +1). \quad (22)$$

$$Pr(r_{2k-i}|d_{2k-i} = -1) = Pr(r_{2k-i}|d_{2k-i} = -1, \hat{d}_{2k-i} = -1) Pr(\hat{d}_{2k-i} = -1|d_{2k-i} = -1)$$
$$+ Pr(r_{2k-i}|d_{2k-i} = -1, \hat{d}_{2k-i} = +1) Pr(\hat{d}_{2k-i} = +1|d_{2k-i} = -1). \quad (23)$$

By defining $P_c$ and $P_e$ as the probability of correct and incorrect symbol detection in the partner side, respectively, we have,

$$P_c = Pr(\hat{d}_{2k-i} = 1|d_{2k-i} = 1) = Pr(\hat{d}_{2k-i} = -1|d_{2k-i} = -1), \quad (24)$$

$$P_e = Pr(\hat{d}_{2k-i} = 1|d_{2k-i} = -1) = Pr(\hat{d}_{2k-i} = -1|d_{2k-i} = 1) = 1 - P_c. \quad (25)$$

Consequently by applying (24) and (25) in (22) and (23) the modified LLR is calculated as follows,

$$y_{2k-i} = \frac{exp\left(-\frac{1}{2\sigma_{eq}^2}\|\mathbf{r}_{2k-i} - \mathbf{h}_{u,k}\mathbf{s}_1 - \mathbf{h}_{p,k}\mathbf{s}_2\|\right)P_c + exp\left(-\frac{1}{2\sigma_{eq}^2}\|\mathbf{r}_{2k-i} - \mathbf{h}_{u,k}\mathbf{s}_1 + \mathbf{h}_{p,k}\mathbf{s}_2\|\right)P_e}{exp\left(-\frac{1}{2\sigma_{eq}^2}\|\mathbf{r}_{2k-i} + \mathbf{h}_{u,k}\mathbf{s}_1 - \mathbf{h}_{p,k}\mathbf{s}_2\|\right)P_e + exp\left(-\frac{1}{2\sigma_{eq}^2}\|\mathbf{r}_{2k-i} + \mathbf{h}_{u,k}\mathbf{s}_1 + \mathbf{h}_{p,k}\mathbf{s}_2\|\right)P_c}. \quad (26)$$

The main problem of the proposed algorithm is that it should know the value of $P_e$ or equivalently $P_c$ which can be variable dependent to the quality of the user partner link. This problem can be solved by considering a pre-determined decision error probability in the partner side. As it can be seen in the next Section, by considering a fixed value for the $P_e$ can noticeably improve the performance of the conventional LLR method in a wide range of channel conditions.

## IV. SIMULATION RESULTS

The proposed structure is simulated by Monte Carlo simulation technique. In this paper, a punctured turbo encoder with 4 states recursive convolutional encoders with generator connection matrix [1 0 1; 1 1 1] is used. Each user and its partner use different spreading sequences. Random codes are considered as spreading sequences. 50 chips random sequences are an efficient model for the long codes used in 3G mobile communications. On the other hand, by the use of random spreading the number of codes can be much more than the spreading length. The number of antenna in all users is assumed to be 1 and BS is assumed to be supported by a 3-element array. The $P_e$ used in all the simulations is experimentally assumed to be 0.025 as it responded well in all the simulations made. This system is simulated in the multiuser case with 50 and 100 interfering users, of course half of the numbers is dedicated to partners, consequently 50 and 100 interfering users correspond to half and full rank CDMA systems. The maximum number considered for the interfering users is twice than the spreading length which is equivalent to full load condition. All users are assumed to be uniformly distributed inside of a circle, and power control is assumed to be used in BS to compensate for the slow fading observed in all user-BS links. Therefore, as introduced in the channel model Section, the averaged received power from each user in BS can be assumed to be unity. The exact modeling of the user-partner links is very complicated, but in this paper we assume they are Ricean channels with log-normal distributions with $R$=5 and 10dB variances dependent on the density of the buildings and other obstacles in the propagation environment. All users are assumed to have random displacement velocities and corresponding $f_D T$ s are considered to have uniform distribution in [0 0.04]. Of course, it must be noted that uniform distribution for $f_D T$ s corresponds to the uniform distribution for users' velocities.

The simulation results are presented in Figs. 6 to 9 for which Figs. 6-7 correspond to when the partner selection set is confined to 10 users, and Figs. 8-9 to an unlimited partner selection set size. Also Figs 6 and 8 are for a half load system which Figs. 7 and 9 are for a full load channel. In all cases, use of cooperation provides noticeable gains over non-cooperation case with much lower BER floors. In all simulation with higher value of the Ricean factor variance, i.e. when $R$=10dB, much better performances can be achieved. This is due to the better quality of the best partner selected among the members in the partner selection set. In other words, in this case, partner can be selected from a wider range of partners Ricean factors. Also, equivalently with a larger partner selection set size i.e. when there is not any limiting assumption on the partner selection set size much better performances can be achieved.

Also, in all cases, out of the 3 combining schemes, the proposed algorithm i.e. the modified LLR based combining presents the best performance and the MRC algorithm presents the worst performance. In BER=$10^{-3}$, the additional performance gain achieved by the modified LLR as compared to the conventional LLR varies in range of 0.2dB, in Fig 6 when R=5dB, to 0.6dB, in Fig 9 when R=10dB.

## V. CONCLUSION

In this paper, performance of the turbo coded Code Division Multiple Access (CDMA) based on cooperation between users in the presence of symbol by symbol time varying Rayleigh fades is evaluated. Also, a new modified LLR based combining scheme is presented to combine the signals received from user and partner side. This algorithm is compared to the MRC and the conventional LLR based algorithms. The system is simulated for a punctured turbo coded structure with 4 states recursive convolutional encoder with generator connection matrix [5 7]. The data symbols are spreaded by 50 chips random spreading sequences after encoding. In multi-user operation, the maximum of 100 interfering users are considered corresponds to a 100 percent load. All users are assumed to be uniformly distributed inside of a circle and the power control is assumed to be used in the BS to compensate for the slow fading observed in all user-BS links. In all simulations, Rayleigh channels are considered for both user-receiver and partner-receiver links and the inter-user links are considered to be Ricean with different Ricean factors with log-normal distribution. All links are considered as fast fading i.e. their coefficients vary symbol by symbol in the duration of a data block.

It is shown that in all cases, use of cooperation provides noticeably high gains over non-cooperation cases. Also, it is shown that in all simulations with higher value of the Ricean factor variance or a larger partner selection set size, much better performances can be achieved. This is due to the better quality of the best partner selected among the members of the partner selection set. Also in all cases, out of the 3 considered combining schemes, the proposed algorithm presents the best performance while the MRC algorithm presents the worst performance.

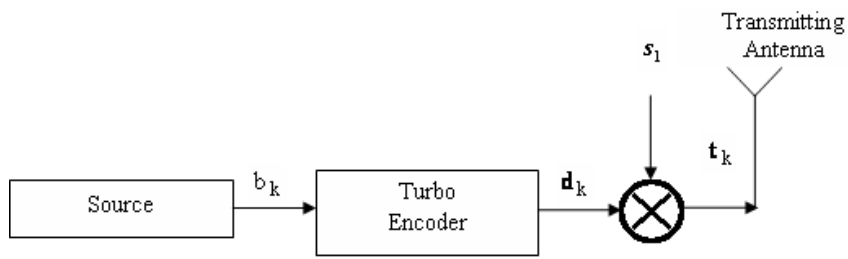

Fig. 1. The block diagram of the transmitter

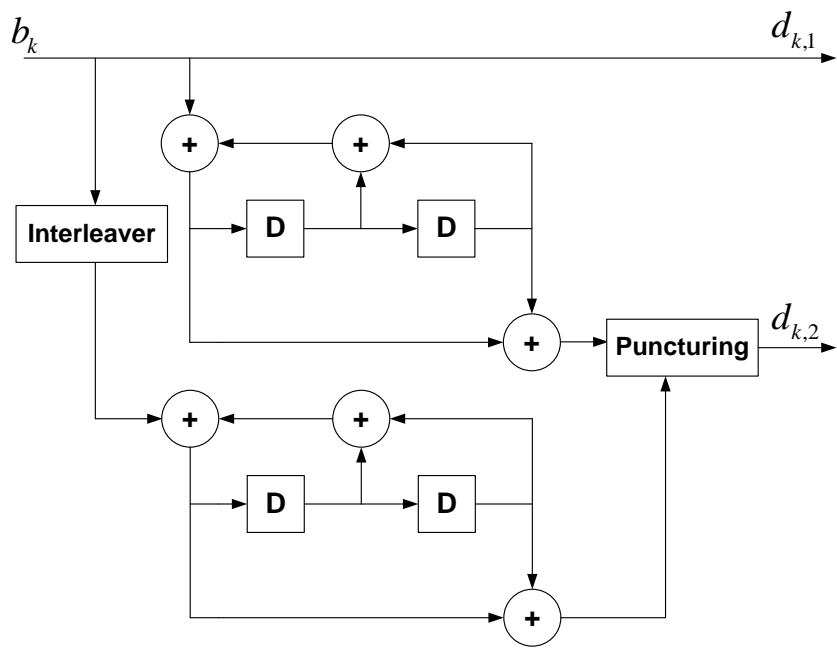

Fig. 2. Block diagram of turbo encoder

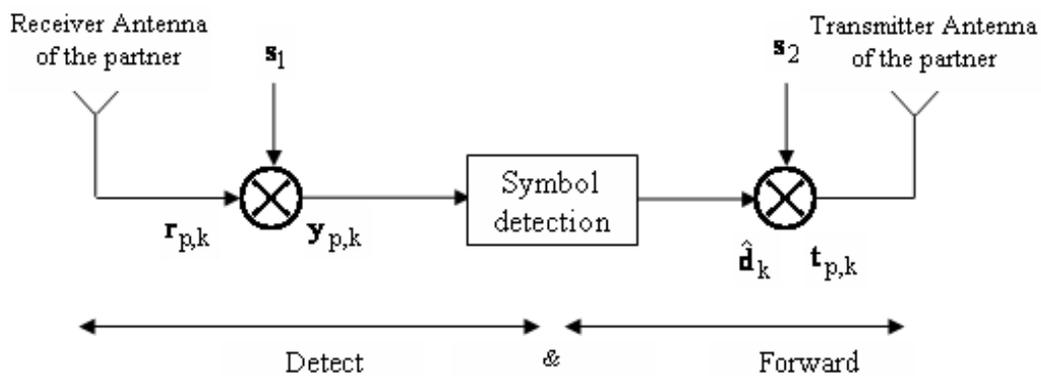

Fig. 3. The partner block diagram

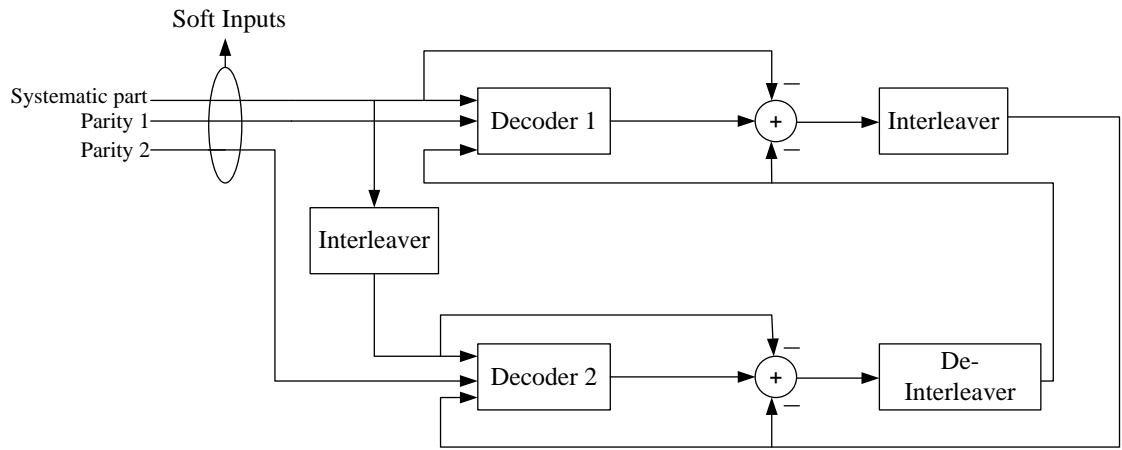

Fig. 4. Turbo decoder block diagram

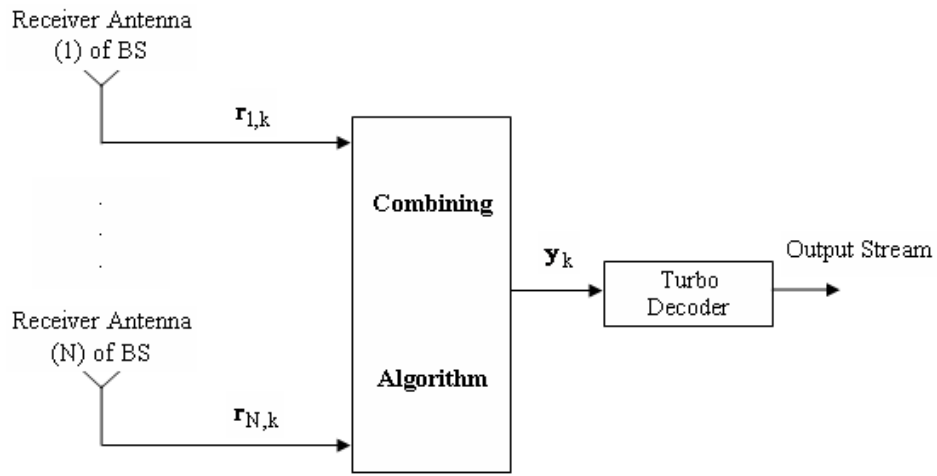

Fig. 5. The BS receiver block diagram

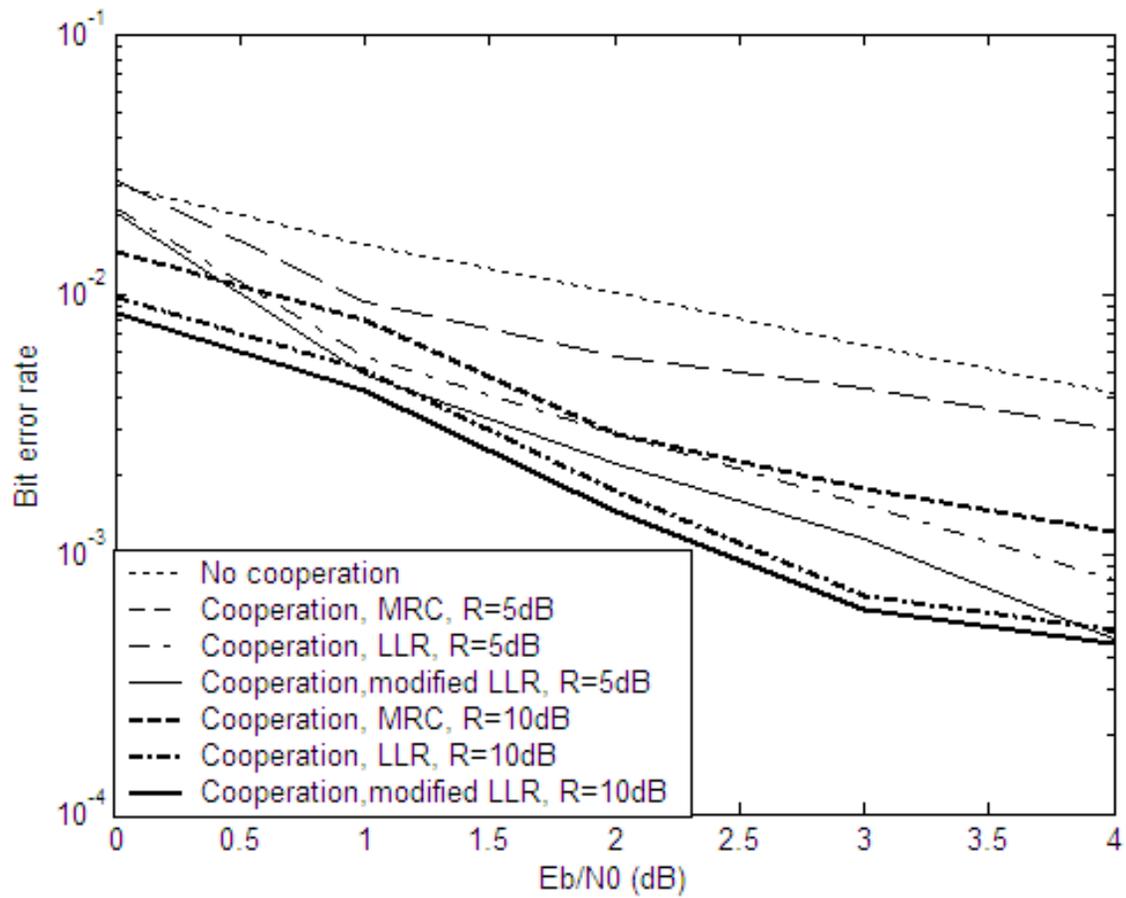

Fig. 6. Comparison of the cooperative system with different combining algorithms for 50 active users and 10 members in partner sets.

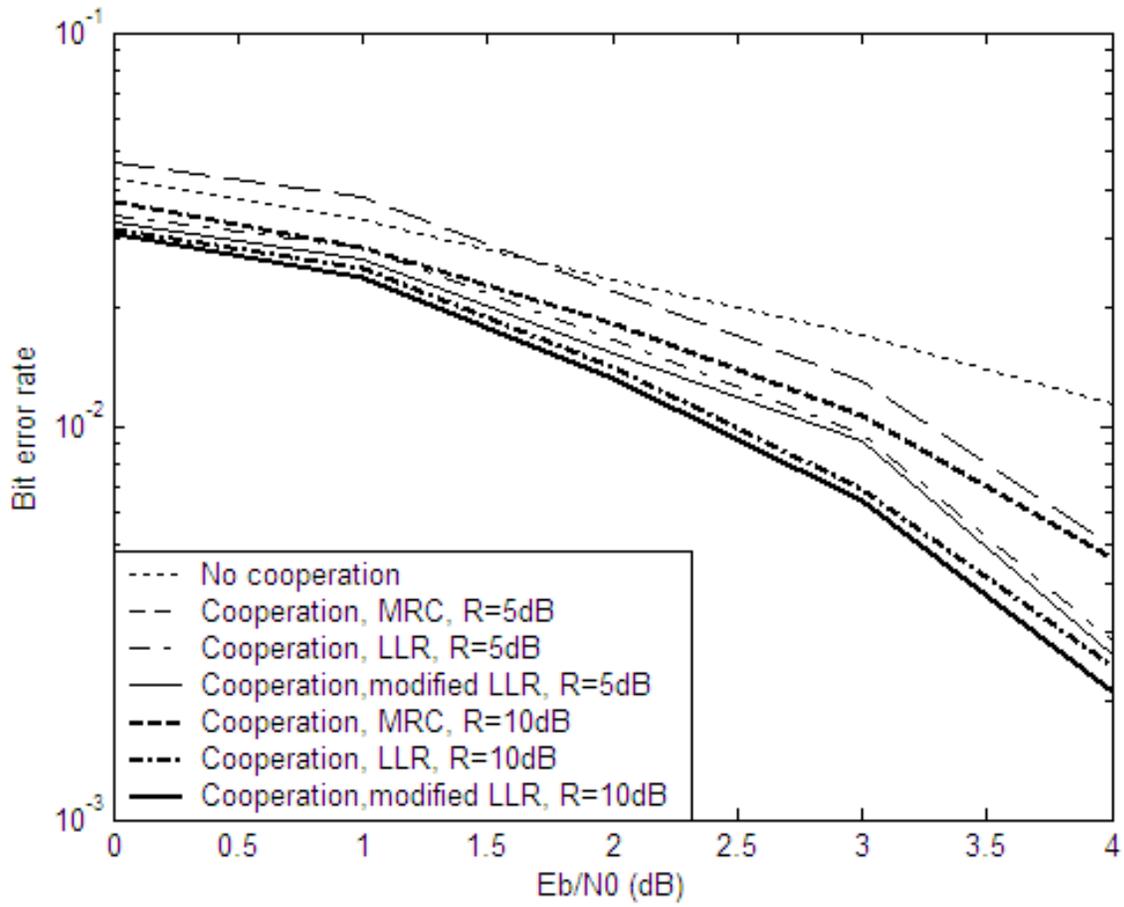

Fig. 7. Comparison of the cooperative system with different combining algorithms for 100 active users and 10 members in partner sets.

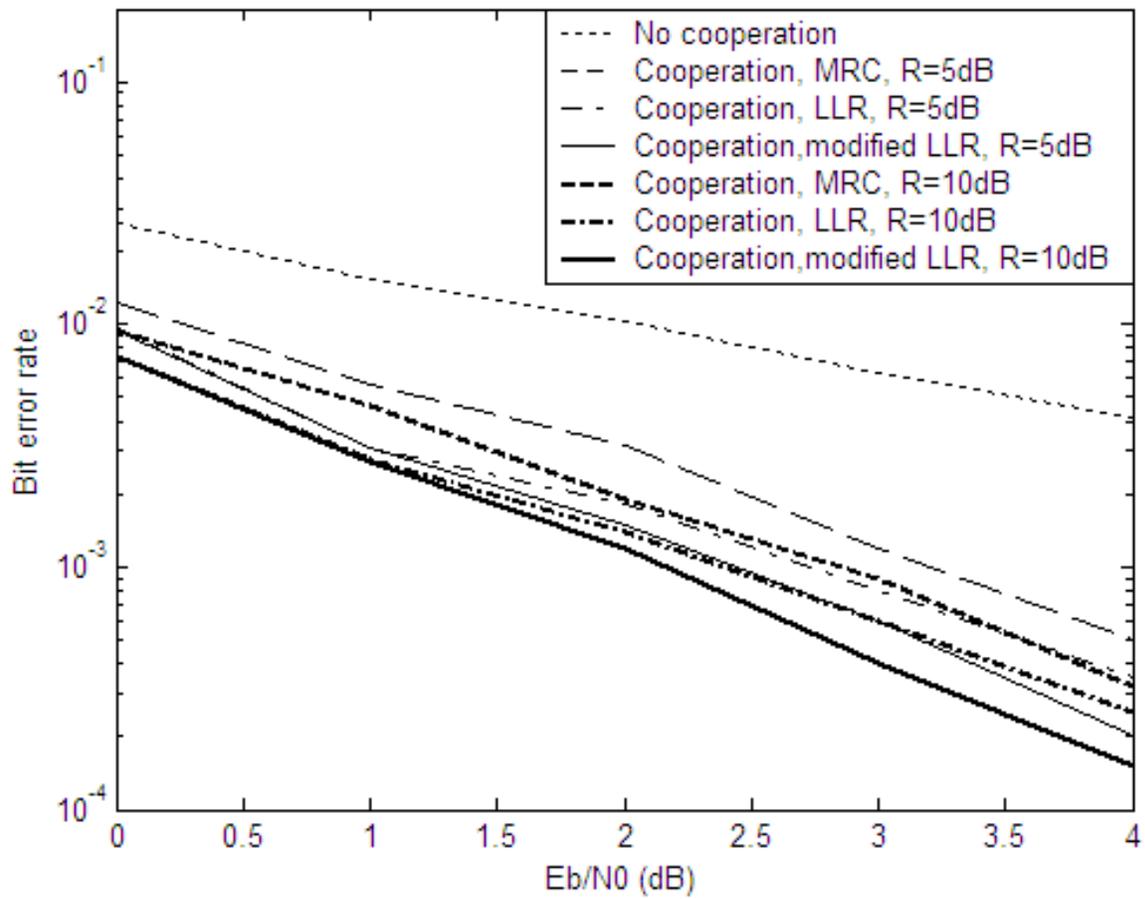

Fig. 8. Comparison of the cooperative system with different combining algorithms for 50 active users and no limitation on partner selection set.

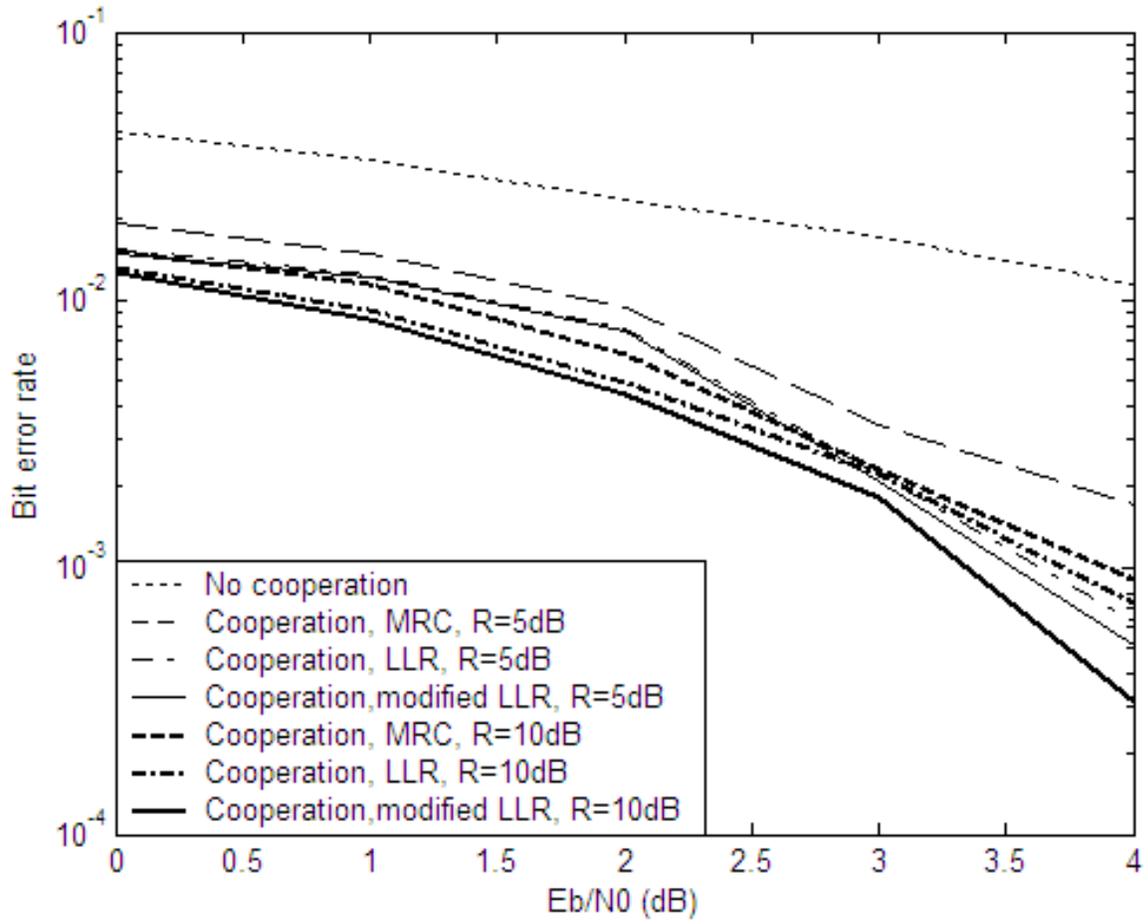

Fig. 9. Comparison of the cooperative system with different combining algorithms for 100 active users and no limitation on partner selection set.